\documentclass[11pt]{article}

\def\be{\begin{equation}}
\def\ee{\end{equation}}
\def\bea{\begin{eqnarray}}
\def\eea{\end{eqnarray}}

\def\ell{l}

\title{Supersymmetric multi-charge $AdS_5$ black holes}

\author{Hari K. Kunduri \\ DAMTP, University of Cambridge, Wilberforce Road, Cambridge, CB3 0WA, UK \\ h.k.kunduri@damtp.cam.ac.uk \\ \\ James Lucietti \\ St. John's College, University of Oxford, Oxford, OX1 3JP, UK \\ j.lucietti@damtp.cam.ac.uk \\  \\Harvey S. Reall \\ School of Physics and Astronomy, University of Nottingham, NG7 2RD, UK \\harvey.reall@nottingham.ac.uk \\ \\ DAMTP-2006-10}
\date{January 20, 2006}

\begin{document}

\maketitle

\begin{abstract}
\noindent A new supersymmetric, asymptotically anti-de Sitter, black hole solution of five-dimensional $U(1)^3$ gauged supergravity is presented. All known examples of such black holes arise as special cases of this solution, which is characterized by three charges and two angular momenta, with one constraint relating these five quantities. Analagous solutions of $U(1)^n$ gauged supergravity are also presented.
\end{abstract}

\section{Introduction}

Supersymmetric black hole solutions of five-dimensional gauged supergravity are of interest because their existence raises the question of whether it is possible to provide an exact calculation of black hole entropy using four dimensional super Yang-Mills theory via the AdS/CFT correspondence \cite{maldacena,gkp,witten}. The first examples of such solutions were obtained two years ago \cite{gr,gr2} and further examples have been obtained more recently \cite{CCLP2,CCLP}. The purpose of this paper is to present a more general supersymmetric solution that contains all of these solutions as special limiting cases.

We shall be concerned with ${\cal N}=1$ gauged supergravity coupled to two abelian vector multiplets, which has gauge group $U(1)^3$. A solution of this theory with (electric) charges $Q_i$ with respect to the three $U(1)$'s can be oxidized to a solution of type IIB supergravity which, in CFT language has $SU(4)$ charge given by the weight vector $(Q_1,Q_2,Q_3)$ \cite{11authors}. Sometimes we shall also refer to minimal gauged supergravity. A charge $Q$ solution of this theory oxidizes to a IIB solution with R-charge $(Q,Q,Q)$ \cite{chamblin}.

The known supersymmetric black holes in five-dimensional gauged supergravity all have non-vanishing angular momentum. Let $J_1$, $J_2$ denote the angular momenta\footnote{$(J_1,J_2)$ is proportional to a weight vector of the rotation group $SO(4)$. In CFT language, it is more common to work with $J_L \equiv J_1 + J_2$ and $J_R \equiv J_1 - J_2$, which are proportional to weights with respect to the two $SU(2)$ factors in $SO(4) \sim SU(2)_L \times SU(2)_R$.}.
The mass is determined by the BPS relation
\be
\label{eqn:bps}
 M = g|J_1| + g|J_2|+ |Q_1| + |Q_2| + |Q_3|,
\ee
where the gauge coupling $g$ is the reciprocal of the AdS radius of curvature.

The solutions of \cite{gr,gr2,CCLP2,CCLP} fall into several
overlapping but distinct families. The 1-parameter solution of
\cite{gr} concerns minimal supergravity, and is a special case of
the 3-parameter $U(1)^3$ solution of \cite{gr2}. The latter is
parameterized by its charges $Q_i$ and has ``self-dual" angular
momenta $J_1 = J_2$. Supersymmetric black hole solutions with $J_1
\ne J_2$ were obtained in \cite{CCLP} and \cite{CCLP2} for the
minimal and $U(1)^3$ theories respectively. The former is not a
special case of the latter. These are both 2-parameter solutions:
the two parameters can be taken to be $J_1$ and $J_2$, which
determine the charges. We have, then, three distinct solutions,
namely those of \cite{gr2,CCLP2,CCLP}. In this paper, we shall
present a more general 4-parameter solution that contains all of
these solutions as special cases. Just like the known solutions, our
solution preserves $1/4$ of the supersymmetry in five dimensions,
which corresponds to $1/16$ supersymmetry (i.e. $2$ real
supercharges) in ten dimensions.

The method that we shall use to construct the solution is based on
the work of \cite{gg,gr2,gs}, which reveals that supersymmetric
solutions of five-dimensional ${\cal N}=1$ gauged supergravity
coupled to abelian vector multiplets can always be written in a
certain canonical form. This canonical form involves a
four-dimensional K\"ahler ``base space". The choice of the base
space is the difficult step in constructing interesting solutions.
The solutions of \cite{gr,gr2} were discovered by requiring the base
space to have a certain highly symmetric form (cohomogeneity 1 with
$U(1) \times SU(2)$ isometry group) which is the reason why these
solutions have $J_1 = J_2$. It turns out that the base space is
singular. One would therefore expect that solutions with $J_1 \ne
J_2$ should also have a singular base space, with less symmetry.
Finding the appropriate space by guesswork would be very difficult.

Fortunately, guesswork is not required because we already know of solutions with $J_1 \ne J_2$, namely those of \cite{CCLP2,CCLP}. These were obtained as limits of non-supersymmetric solutions rather than by using the approach of \cite{gg,gr2,gs}. So we start by writing these solutions in the canonical form. We find that both of these solutions have the same base space, which is indeed a less symmetric generalization (cohomogeneity 2 with $U(1)^2$ isometry group) of that of \cite{gr,gr2}. In other words, we have a base space that encompasses all of the known solutions. We take this base space as the starting point in our generalization of these solutions. Applying the method of \cite{gg,gr2,gs} together with a little more guesswork based on the known solutions then leads to our new solution.

Rather than working directly with the $U(1)^3$ theory, we have found it more convenient to study gauged supergravity coupled to $n-1$ abelian vector multiplets (i.e. gauge group $U(1)^n$) since this was the theory considered in \cite{gr2,gs}. In section 2, we shall obtain a $(n+1)$-parameter supersymmetric black hole solution of this more general theory, and then specialize to the $U(1)^3$ theory in section 3. We have tried to make section 3 self-contained so that it can be read independently of section 2.

The supersymmetric solution that we present should arise as a limit of a non-supersymmetric stationary black hole solution. It would be interesting to find this solution, which should generalize the non-supersymmetric solutions in \cite{CCLP2,CCLP}. A possible method for systematically attacking this problem would be to use the Carter separability criterion, which is equivalent to the existence of a second rank Killing tensor. Indeed, this is how the four dimensional Kerr-Newmann AdS solution was found \cite{carter} and the non-supersymmetric solution of \cite{CCLP} also satisfies this criterion \cite{DKL}.

Finally, we should note that there have been attempts to calculate the entropy of supersymmetric, asymptotically anti-de Sitter black holes using ${\cal N}=4$ $SU(N)$ super Yang-Mills theory. Free field theory can reproduce certain qualitative aspects of these black holes for large R-charge, but, as one might expect, always gives an overestimate of the entropy \cite{rr,kmm}. An alternative approach is to construct an index which only receives contributions from states in short superconformal multiplets that cannot combine into long ones. All such indices where constructed in \cite{kmm}. Unfortunately, these indices count bosonic and fermionic states with opposite signs, which leads to a dramatic cancellation: the index is ${\cal O}(1)$ at large $N$ whereas the black hole entropy is ${\cal O}(N^2)$ \cite{kmm}. The authors of \cite{kmm} speculated that other approaches, taking more account of the dynamics of the gauge theory (e.g. by studying the chiral ring) might be more fruitful. We hope that our work will lead to renewed interest in this problem.

\section{The general solution}

\subsection{The theory}

\label{sec:theory}

We shall consider the theory of five dimensional ${\cal N}=1$ gauged supergravity coupled to $n-1$ abelian vector multiplets constructed in \cite{gst}. The bosonic sector this theory consists of the graviton, $n$ vectors $A^I$ and $n-1$ real scalars. The latter can be replaced with $n$ real scalars $X^I$ subject to a constraint
\be
 \frac{1}{6}C_{IJK} X^I X^J X^K = 1,
\ee
where $C_{IJK}$ are a set of real constants symmetric under permutations of $(IJK)$. Indices $I,J,K, \ldots$ run from $1$ to $n$. It is convenient to define
\be
\label{eqn:Xconstr}
 X_I \equiv \frac{1}{6} C_{IJK} X^J X^K.
\ee
The action is\footnote{We use a positive signature metric.}
\be
 S = \frac{1}{16 \pi G} \int \left( R_5 \star 1  - Q_{IJ} F^I \wedge \star F^J - Q_{IJ}     dX^I \wedge \star dX^J - \frac{1}{6} C_{IJK} F^I \wedge F^J \wedge A^K  + 2g^2 {\cal V} \star 1 \right),
\ee
where $F^I \equiv dA^I$ and
\be
 Q_{IJ} \equiv \frac{9}{2} X_I X_J - \frac{1}{2} C_{IJK} X^K.
\ee
For simplicity, we shall assume that the scalars parametrize a symmetric space, which is equivalent to the condition
\be
 C_{IJK} C_{J' (LM} C_{PQ) K'} \delta^{JJ'} \delta^{KK'} = \frac{4}{3} \delta_{I(L} C_{MPQ)}.
\ee
This condition ensures that the matrix $Q_{IJ}$ is invertible, with inverse
\be
 Q^{IJ} = 2 X^I X^J - 6 C^{IJK} X_K,
\ee
where
\be
 C^{IJK} \equiv C_{IJK}.
\ee
We then have
\be
 X^I = \frac{9}{2} C^{IJK} X_J X_K.
\ee
The symmetric space condition is probably not essential -- we expect that it should be possible to use the results of \cite{gs} to relax it in what follows. However, we are mainly interested in a $U(1)^3$ theory for which this condition is satisfied so we shall assume it for simplicity henceforth.

The scalar potential is\footnote{This is related to the notation of \cite{gr2} via equations (2.23) and (2.29) of that paper.}
\be
 {\cal V} = 27 C^{IJK} \bar{X}_I \bar{X}_J X_K,
\ee
where $\bar{X}_I$ are a set of constants. It was shown in \cite{gr2} that the unique maximally supersymmetric solution of this theory is $AdS_5$ with radius $g^{-1}$, vanishing vectors and constant scalars: $X^I = \bar{X}^I$, where
\be
 \bar{X}^I \equiv \frac{9}{2} C^{IJK} \bar{X}_J \bar{X}_K.
\ee
In section 3, we shall consider a particular $U(1)^3$ gauged supergravity. In the above language, this theory has $n=3$, $C_{IJK} = 1$ if $(IJK)$ is a permutation of $(123)$ and $C_{IJK} = 0$ otherwise, and $\bar{X}^I=1$ (so $\bar{X}_I=1/3$).

\subsection{Supersymmetric solutions}

The general nature of supersymmetric solutions of this theory was
deduced in \cite{gr2} following closely the corresponding analysis
for the minimal theory given in \cite{gg}. Given a supercovariantly
constant spinor $\epsilon$, one can construct a real scalar $f \sim
\bar{\epsilon} \epsilon$ and a real vector $V^\mu \sim
\bar{\epsilon} \gamma^\mu \epsilon$. These obey $V^2 = -f^2$, so $V$
is timelike or null, and it turns out that $V$ is always Killing.
There are two cases: a ``null" case, in which $V$ is globally null
and a ``timelike" case in which $V$ is timelike in some region
${\cal U}$ of spacetime. The former case was treated in \cite{gg,gs}
and does not concern us here because such solutions cannot describe
black holes.

In the timelike case, we can, without loss of generality, assume that $f>0$ in ${\cal U}$, and introduce local coordinates so that the metric takes the form
\be
\label{eqn:metric}
 ds^2 = -f^2 \left( dt + \omega \right)^2 + f^{-1} h_{mn} dx^m dx^n,
\ee with $V = \partial/\partial t$, $h_{mn}$ is a metric on a
4-dimensional Riemannian ``base space" $B$ and $\omega$ a 1-form on
$B$. Supersymmetry implies that the base space is K\"ahler
\cite{gg,gr2}. Let $J$ denote the K\"ahler form and define the
orientation of $B$ so that $J$ is anti-self-dual. If $\eta_4$
denotes the volume form of $B$ then $(dt + \omega) \wedge \eta_4$
must then be positively oriented in space-time.

Turning to the Maxwell fields, supersymmetry implies that \cite{gr2}
\be
\label{eqn:maxwell}
 F^I = d \left[ X^I f (dt + \omega) \right] + \Theta^I - 9 g f^{-1} C^{IJK} \bar{X}_J X_K J,
\ee
and
\be
\label{eqn:thetaconstr}
 X_I \Theta^I = - \frac{2}{3} G^+,
\ee
where $\Theta^I$ are self-dual 2-forms on $B$ and
\be
 G^+ = \frac{1}{2} f \left( d\omega + \star_4 d\omega \right),
\ee
where $\star_4$ is the Hodge dual on $B$. Now define ${\cal R}$ to be the Ricci form of $B$:
\be
 {\cal R}_{mn} \equiv \frac{1}{2} R_{mnpq} J^{pq},
\ee
where $R_{mnpq}$ is the Riemann tensor of $B$ and indices are raised with $h^{mn}$. Supersymmetry implies that \cite{gr2}
\be
\label{eqn:kahlerpot}
{\cal R} = d P,
\ee
where
\be
 P_m \equiv 3g \bar{X}_I \left( A^I_m - f X^I \omega_m \right).
\ee
These equations determine $f$ \cite{gr2} ($R$ is the Ricci scalar of $B$):
\be
\label{eqn:fsol}
 f = -\frac{108g^2}{R} C^{IJK} \bar{X}_I \bar{X}_J X_K.
\ee
The above conditions are all necessary for supersymmetry. The analysis of \cite{gr2} reveals that they are also {\it sufficient}, i.e., a supercovariantly constant spinor will exist if the metric and maxwell field are given by (\ref{eqn:metric}) and (\ref{eqn:maxwell}) for some K\"ahler $B$, and equations (\ref{eqn:thetaconstr}) and (\ref{eqn:kahlerpot}) are satisfied. The field equations of the theory are all satisfied once we impose the equations of motion for the Maxwell fields \cite{gr2}, i.e., the Bianchi identities
\be
 dF^I = 0,
\ee
and the Maxwell equations,
\be
 d \left( Q_{IJ} \star F^J \right) = - \frac{1}{4} C_{IJK} F^J \wedge F^K.
\ee
One can substitute the expression (\ref{eqn:maxwell}) into these equations to obtain a pair of equations on $B$ \cite{gr2} but we shall not write them out here.

The analysis of \cite{gg,gr2} reveals that the fraction of supersymmetry preserved by these solutions is generically $1/4$, i.e., such solutions are preserved by $2$ real supercharges. This might be enhanced in special cases. However, it has been shown \cite{ggn} that this does not happen for the supersymmetric black holes of \cite{gr}, even when oxidized to ten dimensions. Since these are a special case of our solution, we do not expect supersymmetry enhancement here either.

\subsection{The base space}

\label{sec:base}

The first step in constructing a new solution using the above procedure is to choose the base space. As discussed in the introduction, we shall use a base space obtained by writing the metrics of the supersymmetric solutions of \cite{CCLP2,CCLP} in the form (\ref{eqn:metric}). We find that the base space of both metrics takes the form
\bea
\label{eqn:basemetric}
 h_{mn} dx^m dx^n &=& (r^2-r_0^2) \left\{
\frac{dr^2}{\Delta_r} +\frac{d\theta^2}{\Delta_{\theta}} +
\frac{\cos^2\theta}{\Xi_b^2} \left[ \Xi_b + \cos^2 \theta \left(\rho^2 g^2 + 2(1+bg)(a+b)g \right) \right]
d\psi^2 \right. \nonumber \\ &+&
\frac{\sin^2\theta}{\Xi_a^2}\left[
\Xi_a + \sin^2 \theta \left(\rho^2 g^2 + 2(1+ag)(a+b) g \right)
\right]d\phi^2 \nonumber \\  &&\left. +
\frac{2\sin^2\theta \cos^2\theta}{\Xi_a\Xi_b}\left[ \rho^2 g^2
+2(a+b) g+(a+b)^2g^2\right]d\psi d\phi \right\},
\eea
where
\bea
 \Delta_r &=& (r^2-r_0^2)^2[g^2r^2+(1+ag+bg)^2]/r^2, \qquad \Delta_\theta = 1 - a^2 g^2 \cos^2\theta - b^2 g^2
\sin^2\theta, \nonumber \\
 \Xi_a &=& 1-a^2 g^2, \qquad \Xi_b = 1-b^2 g^2, \qquad \rho^2 = r^2 + a^2 \cos^2\theta + b^2 \sin^2\theta.
\eea
The coordinate ranges are $r^2>r_0^2$, $0 \le \theta \le \pi/2$ and $\phi,\psi$ are angular coordinates with period $2\pi$. Surfaces of constant $r$ have topology $S^3$.
This metric is specified by three constants: $a$, $b$ and $r_0^2$. The first two of these correspond roughly to angular momentum parameters for the $\phi$ and $\psi$ directions of the black holes of \cite{CCLP2,CCLP}. We shall assume\footnote{
It is easy to see that $a,b \ge 0$ can always be arranged by a coordinate transformation. For example, if $a<0$ in the general non-supersymmetric solution of \cite{CCLP} then taking $t \rightarrow -t$, $\psi \rightarrow -\psi$ and $q \rightarrow -q$ (where $q$ is the charge parameter of \cite{CCLP}) effectively reverses the sign of $a$. The base space is determined {\it after} arranging $a\ge 0,b\ge 0$.
}
$g^{-1} > a,b \ge 0$. The base space has a curvature singularity at $r^2=r_0^2$ but the space-time metric (\ref{eqn:metric}) is smooth at $r^2=r_0^2$, in fact this corresponds to the event horizon. $r_0^2$ is not an independent parameter for these solutions. For the minimal gauged supergravity solution of \cite{CCLP}, it takes the value $r_0^2 = r_m^2$, where
\be
\label{eqn:rmdef}
 r_m^2 = g^{-1} (a + b + a b g).
\ee
For the $U(1)^3$ solution of \cite{CCLP2}, $r_0^2 = r_*^2$, where
\be
 r_*^2 = \frac{ab}{1+ ag + bg}.
\ee
For these values of $r_0$, supersymmetry implies that the above base space metric must be K\"ahler. We find that the K\"ahler form is
\be
\label{eqn:kahlerform}
 J = -d \left[ \frac{(r^2-r_0^2)}{2} \left( \frac{\cos^2 \theta d\psi}{\Xi_b} + \frac{\sin^2 \theta d\phi}{\Xi_a} \right) \right].
\ee
In fact, we find that the metric (\ref{eqn:basemetric}) is K\"ahler for {\it any} $r_0^2$, with K\"ahler form given by (\ref{eqn:kahlerform}). Anti-self-duality of $J$ determines the orientation of $B$: $dr \wedge d\theta \wedge d\psi \wedge d\phi$ is positively oriented. For later convenience, we also record the Ricci form:
\be
\label{eqn:ricciform}
 {\cal R} = 3g^2 d \left[ \frac{\cos^2 \theta}{\Xi_b} \left(\rho^2 + b^2 + \frac{4 r_m^2}{3} - \frac{r_0^2}{3} \right) d\psi + \frac{\sin^2 \theta}{\Xi_a} \left( \rho^2 + a^2 + \frac{4 r_m^2}{3} - \frac{r_0^2}{3} \right) d\phi \right].
\ee
It appears, then, that we have a 3-parameter generalization of the 2-parameter base metrics of \cite{CCLP2,CCLP}. However, this is not the case: it turns out one combination of these parameters can be eliminated by a coordinate transformation. To see this, let
\be
\label{eqn:sigmadef}
 r^2-r_0^2=\alpha^2 \sinh^2(g\sigma), \qquad
 \alpha^2  \equiv r_0^2 +g^{-2}(1+ag+bg)^2.
\ee
The base metric now takes the simplified form\footnote{
In this form, it is easy to see that the base metric reduces to the cohomogeneity 1 base metric of \cite{gr,gr2} when $a=b$.}:
\bea
\label{eqn:simplebase}
\nonumber h_{mn}dx^mdx^n &=& d\sigma^2 + \alpha^2\sinh^2(g\sigma) \left( \alpha^2 g^2
\cosh^2(g\sigma) -\Delta_{\theta} \right) \left(\cos^2\theta
\frac{d\psi}{\Xi_b} + \sin^2\theta \frac{d\phi}{\Xi_a}\right)^2
\\ &+& \alpha^2\sinh^2(g\sigma) \left( \frac{d\theta^2}{\Delta_{\theta}} +
\Delta_{\theta}\cos^2\theta \frac{d\psi^2}{\Xi_b^2} +
\Delta_{\theta}\sin^2\theta \frac{d\phi^2}{\Xi_a^2} \right).
\eea
Since $\Delta_{\theta}=\Xi_a \cos^2\theta+ \Xi_b \sin^2\theta$, it can now be seen
that the metric depends on only two combinations of the constants, namely $A,B>0$, where
\be
\label{AB}
 A^2 \equiv \frac{\Xi_a}{g^2 \alpha^2}, \qquad B^2 \equiv \frac{\Xi_b}{g^2 \alpha^2}.
\ee The redundancy in the metric (\ref{eqn:basemetric}) implies that
$r_0^2$ can be set to any convenient value by means of a coordinate
transformation. We shall choose $r_0^2 = 0$. Note that the
parameters $a$, $b$ also transform. If we start in the ``gauge"
$r_0=0$ and transform to some new value of $r_0$ then the new values
$(a',b')$ can be determined by invariance of $A,B$. This is
important when comparing our results with those of
\cite{CCLP2,CCLP}.

\subsection{The solution}

Having chosen a base space, it remains to solve the remaining conditions for supersymmetry. We shall do this by making an Ansatz that is sufficiently general to encompass the known solutions of \cite{gr2,CCLP2,CCLP}. We take the general supersymmetric metric (\ref{eqn:metric}) with the particular base space (\ref{eqn:basemetric}) with $r_0=0$. The Ansatz is
\be
 \omega = \omega_\psi d\psi + \omega_\phi d\phi,
\ee
\be
 \omega_\psi = - \frac{g \cos^2 \theta}{r^2 \Xi_b} \left( \rho^4 + P_2 \rho^2 + P_0 \right), \qquad \omega_\phi = - \frac{g \sin^2 \theta}{r^2 \Xi_a} \left( \rho^4 + Q_2 \rho^2 + Q_0 \right),
\ee
\be
\label{eqn:Xansatz}
 f^{-1} X_I = \frac{1}{3} H_I \equiv \frac{\bar{X}_I \rho^2 + e_I}{r^2},
\ee
\be
 A^I = fX^I (dt + \omega) +  \frac{g \cos^2 \theta}{\Xi_b} \left( \bar{X}^I \rho^2 + c^I \right) d\psi + \frac{g \sin^2 \theta}{\Xi_a} \left( \bar{X}^I \rho^2 + d^I  \right) d\phi,
\ee
where $P_2,P_0,Q_2,Q_0,e_I,c^I$ and $d^I$ are constants. Equations (\ref{eqn:Xansatz}) and (\ref{eqn:Xconstr}) determine $f$:
\be
\label{eqn:fH}
 f^{-3} = \frac{1}{6} C^{IJK} H_I H_J H_K,
\ee
i.e.,
\be
 f = \frac{r^2}{F(\rho^2)^{1/3}}, \qquad F(\rho^2) \equiv \rho^6 + \beta_1 \rho^4 + \beta_2 \rho^2 + \beta_3,
\ee
where
\be
 \beta_1 = \frac{27}{2} C^{IJK}\bar{X}_I\bar{X}_Je_K = 3 \bar{X}^I e_I, \qquad \beta_2 = \frac{27}{2} C^{IJK}\bar{X}_I e_J e_K, \qquad \beta_3 = \frac{9}{2}C^{IJK}e_Ie_Je_K.
\ee
For the solutions of \cite{gr2,CCLP2,CCLP}, $A^I_\psi$ and $A^I_\phi$ decay as $1/r^2$ for large $r$ so we impose the same condition here, which determines
\be
 c^I = \bar{X}^I \left(P_2 - \beta_1 \right) + 9 C^{IJK} \bar{X}_J e_K, \qquad
 d^I = \bar{X}^I \left(Q_2 - \beta_1 \right) + 9 C^{IJK} \bar{X}_J e_K.
\ee
For orientation, we note that this Ansatz can be embedded in minimal gauged supergravity when $X_I \equiv \bar{X}_I$ and $A^I = \bar{X}^I {\cal A}$ where ${\cal A}$ is the Maxwell potential of the minimal theory.

Now we impose the conditions required for supersymmetry. First,
equation (\ref{eqn:kahlerpot}) is satisfied for the Ricci form
(\ref{eqn:ricciform}) by taking $P_2 = (4r_m^2+\beta_1)/3 + b^2$ and
$Q_2 = (4r_m^2+\beta_1)/3 + a^2$ (where $r_m^2$ is defined by
(\ref{eqn:rmdef})). Next, either from imposing self-duality of
$\Theta^I$ or from equation (\ref{eqn:fsol}) one obtains the
constraint \be \label{eqn:econstr}
 \beta_1 = 2r_m^2,
\ee
and hence
\be
\label{P2Q2}
 P_2 = 2r_m^2 + b^2, \qquad Q_2 = 2r_m^2 + a^2.
\ee
The final equation required for supersymmetry is (\ref{eqn:thetaconstr}), which is satisfied if
\be
P_0  =  \frac{1}{2} \left[\beta_2-a^2 b^2 + g^{-2} (a^2-b^2)\right], \qquad Q_0  =  \frac{1}{2} \left[\beta_2-a^2 b^2 - g^{-2} (a^2-b^2)\right].
\ee
We must now impose the equations of motion for the Maxwell fields. The Bianchi identities are trivial (we've specified potentials) and the Maxwell equations turn out to impose no further restrictions. Hence we have a supersymmetric solution. It is specified by the constants $a,b$ and $e_I$ subject to the constraint (\ref{eqn:econstr}). Therefore the solution has $n+1$ independent parameters.

\subsection{Asymptotics and charges}
Now we show that our supersymmetric solution asymptotes to $AdS_5$.
To transform to a frame which is not rotating at infinity let
$t=\bar{t}$, $\psi = \bar{\psi}-g \bar{t}$, $\phi =
\bar{\phi}-g\bar{t}$ and $y^2=r^2+ 2 r_{m}^2/3$. The spacetime
metric for large $r$ (or $y$) then takes the form \bea \nonumber
ds^2 &=& -\left[ \frac{\Delta_{\theta}}{\Xi_a\Xi_b}(1+g^2y^2)
+\mathcal{O}\left(\frac{1}{y^2}\right) \right]d\tilde{t}^2 +
\mathcal{O}\left( \frac{1}{y^2} \right) d\tilde{t}d\tilde{\psi}
+\mathcal{O}\left( \frac{1}{y^2}\right) d\tilde{t}d\tilde{\phi} +
\mathcal{O}\left( \frac{1}{y^2} \right) d\tilde{\psi}d\tilde{\phi}
\\ \nonumber &+& \left[ \frac{\cos^2\theta}{\Xi_b}(y^2+b^2)+
\mathcal{O}\left(\frac{1}{y^2}\right) \right]d\tilde{\psi}^2 +\left[
\frac{\sin^2\theta}{\Xi_a}(y^2+a^2)+
\mathcal{O}\left(\frac{1}{y^2}\right) \right]d\tilde{\phi}^2 \\ &+&
\frac{\tilde{\rho}^2+\mathcal{O}(1/y^2)}{\Delta_{\theta}}d\theta^2 +
\frac{\tilde{\rho}^2+\mathcal{O}(1/y^2)}{\Delta_y}dy^2 \\
\nonumber \tilde{\rho}^2 &=&y^2+a^2\cos^2\theta+b^2\sin^2\theta, \\
\nonumber \Delta_y &=& y^{-2}\left(y^2- \frac{2}{3}r_{m}^2 \right)^2
\left( 1+g^2y^2+g^2a^2+g^2b^2+ \frac{4}{3}g^2r_{m}^2 \right).\eea
To bring the asymptotic metric into the
familiar global $AdS_5$ coordinates one needs to perform the
transformation:
\be \Xi_a Y^2\sin^2\Theta = (y^2+a^2)\sin^2\theta,
\qquad \Xi_b Y^2 \cos^2\Theta = (y^2+b^2) \cos^2\theta.
\ee
In the asymptotically static coordinates, the supersymmetric Killing field is
\be
\label{eqn:Vstatic}
 V = \frac{\partial}{\partial \bar{t}} + g \frac{\partial}{\partial \bar{\phi}} + g \frac{\partial}{\partial \bar{\psi}}.
\ee
The notation used above enabled the solution to be presented in a concise manner. However, in calculating the charges it is useful to define new parameters $q_I$ defined by
\be
 e_I = \sqrt{\Xi_a \Xi_b} q_I - g^{-2} \left(1 - \sqrt{\Xi_a \Xi_b}\right) \bar{X}_I.
\ee
We shall see that $q_I$ are closely related to the electric charges of our solution. Let
\be
 \alpha_1 = \frac{27}{2} C^{IJK}\bar{X}_I\bar{X}_Jq_K = 3 \bar{X}^I q_I, \qquad \alpha_2 = \frac{27}{2} C^{IJK}\bar{X}_I q_J q_K, \qquad \alpha_3 = \frac{9}{2}C^{IJK}q_Iq_Jq_K.
\ee
These are related to the $\beta_i$ via
\bea
 \beta_1 &=& \sqrt{\Xi_a\Xi_b}\alpha_1 - 3g^{-2}(1-\sqrt{\Xi_a\Xi_b}), \nonumber \\
 \beta_2 &=& \Xi_a\Xi_b\alpha_2 -\frac{2\sqrt{\Xi_a\Xi_b}\left(1-\sqrt{\Xi_a\Xi_b} \right)}{g^2}\alpha_1 +\frac{3\left(1-\sqrt{\Xi_a\Xi_b}\right)^2}{g^4}, \nonumber \\
\beta_3 &=& (\Xi_a\Xi_b)^{\frac{3}{2}}\alpha_3 -
\frac{\Xi_a\Xi_b}{g^2}\left(1-\sqrt{\Xi_a\Xi_b}\right)\alpha_2
+\frac{\sqrt{\Xi_a\Xi_b}(1-\sqrt{\Xi_a\Xi_b})^2}{g^4}\alpha_1 -
\frac{\left(1-\sqrt{\Xi_a\Xi_b}\right)^3}{g^6}. \eea The
constraint~(\ref{eqn:econstr}) becomes
\be \alpha_1 = \frac{1}{g^2}
\left(\frac{1}{AB} + \frac{A}{B} + \frac{B}{A} -3 \right) =
 \frac{1}{\sqrt{\Xi_a \Xi_b}} \left[ 2r_m^2 + 3 g^{-2} \left( 1 - \sqrt{\Xi_a \Xi_b} \right) \right],
\ee where\footnote{Note that $A$, $B$ are the ``gauge-invariant"
quantities defined in (\ref{AB}), here written in the gauge $r_0=0$.
It is therefore straightforward to convert our expressions to any
other value for $r_0$, which facilitates comparison
with~\cite{CCLP2,CCLP}.} \be
 A \equiv \frac{\sqrt{\Xi_a}}{(1 + ag + bg)}, \qquad B \equiv \frac{\sqrt{\Xi_b}}{(1+ag+bg)}.
\ee
The electric charges are defined by:
\begin{equation}
Q_I = \frac{1}{8\pi G} \int_{S^3}  Q_{IJ} \star F^J,
\end{equation}
where the integral is taken over a three-sphere at
infinity. This gives:
\begin{equation}
Q_I = \frac{3\pi}{4G} \left(q_I - \frac{g^2\alpha_2}{2}\bar{X}_{I} + \frac{3g^2}{2}C_{IJK}\bar{X}^JC^{KLM}q_{L}q_{M} \right ).
\end{equation}
Note that this is independent of $a,b$ and hence coincides with the result of \cite{gr2}.

We will define the angular momenta via Komar integrals
\be
 J_i =
\frac{1}{16\pi G} \int_{S^3} \star dK_i,
\ee
where the Killing vector
$K_i^{\mu}$ is either $\partial_{\phi}$ or $\partial_{\psi}$ and the $S^3$ is at infinity. We find:
\bea
J_{\phi} &=& \frac{\pi}{4G}\left[ \frac{1}{2}g\alpha_2+ g^3\alpha_3
+\frac{(B-A)}{Ag^3}\left( 1+g^2\alpha_1+g^4\alpha_2+g^6\alpha_3 \right) \right],
\\
J_{\psi} &=& \frac{\pi}{4G}\left[ \frac{1}{2}g\alpha_2+ g^3\alpha_3
+\frac{(A-B)}{Bg^3} \left( 1+g^2\alpha_1+g^4\alpha_2+g^6\alpha_3
\right) \right]. \eea
The mass is fixed by the BPS condition \be
 M=g|J_\phi|+g|J_\psi| + |\bar{X}^I Q_I|,
\ee
which gives
\bea
M= \frac{\pi}{4G} \left[ \alpha_1+ \frac{3}{2}g^2\alpha_2 + 2g^4
\alpha_3 +\frac{(A-B)^2}{g^2AB}\left(1+g^2\alpha_1 +g^4 \alpha_2 +g^6
  \alpha_3 \right) \right].
\eea
As written above, is trivial to see that these various charges reduce to
those of the black holes of~\cite{gr2} when $a=b$. We have have also checked that
they are consistent with the results of~\cite{CCLP2,CCLP}, for which
one needs to use (\ref{AB}) together with the appropriate value of $r_0$.

\subsection{Absence of causal pathologies}

Note that $f$ is an invariant of the solution and should therefore remain finite on, and outside, the horizon.
Hence we demand
\be
 F(\rho^2) > 0 \quad {\rm for} \quad r \ge 0.
\ee This constraint guarantees that the scalars are finite for $r
\ge 0$. For $A>B$, evaluating this inequality at $r=0$ gives
\be
g^6\alpha_3 > \frac{(A-B)}{B} \left( \frac{(A-B)^2}{B^2}
+g^4\alpha_2 \right). \ee
If $A<B$ then the same expression holds
with $A$ and $B$ interchanged.

Absence of closed causal curves requires that the $\phi-\psi$ part of the metric be positive definite. We find that this is the case as $r \rightarrow 0$ if, and only if,
\begin{equation}\label{noCTC}
\delta \equiv (1+g^2\alpha_1)\alpha_3 - \frac{g^2\alpha_2^2}{4} - \frac{(A-B)^2}{ABg^6}\left(1+g^2\alpha_{1}+g^4\alpha_2 + g^6\alpha_3\right) > 0.
\end{equation}

\subsection{Extension through the horizon}

We shall now show that our solution has an event horizon at $r=0$. To this end, we transform to new coordinates $(v,R,\theta,\phi',\psi')$ where
\be
 R = gr^2,
\ee
\be
 dv = dt - \left(\frac{A_0}{g^2 R^2} + \frac{A_1}{gR} \right) dR, \qquad d\psi' = d\psi - \frac{B_0 \Xi_b}{R} dR,   \qquad d\phi' = d\phi - \frac{C_0 \Xi_a}{R} dR,
\ee
where $A_0,A_1,B_0,C_0$ are constants to be determined. The exterior of the black hole corresponds to $R>0$. The scalars $X_I$ can be smoothly extended through $R=0$, as can the $v,\phi'$ and $\psi'$ components of the Maxwell potentials $A^I$. The remaining non-zero components $A^I_R$ diverge as $1/R$ as $R \rightarrow 0$. However, this divergence is pure gauge if the coefficient of $1/R$ is independent of $\theta$. This will be the case if, and only if,
\be
\label{eqn:hor1}
 C_0 \left(Q_0 - a^2 Q_2 - \beta_2 \right) = B_0 \left(P_0 - b^2 P_2 - \beta_2 \right),
\ee
\be
\label{eqn:hor2}
 B_0 \left(b^2 P_0 + \beta_3 \right) - C_0 \left(a^2 Q_0 + \beta_3 \right) + g^{-4} (a^2-b^2) A_0 = 0.
\ee
If $A_0,B_0,C_0$ satisfy these equations then the Maxwell field strengths can be smoothly extended through $R=0$.

Now let's consider the metric. It is convenient to work with independent variables $R$ and $\rho^2$ rather than $R$ and $\theta$. We then expand metric components as Laurent series in $R$ (near $R=0$). The coefficients in this series are analytic functions of $\rho^2$. We have
\begin{eqnarray}
 g_{vv} = -f^2 = {\cal O}(R^2), \qquad g_{v\psi'} &=& -f^2 \omega_\psi = {\cal O}(R), \qquad g_{v\phi'} = -f^2 \omega_\phi = {\cal O}(R), \nonumber \\
 g_{\phi'\phi'} = g_{\phi\phi} = {\cal O}(1), \qquad g_{\phi'\psi'} &=& g_{\phi \psi} = {\cal O}(1), \qquad g_{\psi'\psi'} = g_{\psi \psi} = {\cal O}(1).
\end{eqnarray}
Using equations (\ref{eqn:hor1}) and (\ref{eqn:hor2}) we find
\be
 g_{vR} = \left( \frac{B_0 - C_0}{a^2 - b^2} \right) F(\rho^2)^{1/3} + {\cal O}(R), \qquad
 g_{R\psi'} = {\cal O}(1), \qquad g_{R\phi'} = {\cal O}(1).
\ee
The $RR$ component of the metric contains divergent $1/R^2$ and $1/R$ terms so we need the coefficients of these terms to vanish. After using (\ref{eqn:hor1}) and (\ref{eqn:hor2}), the vanishing of the coefficient of the $1/R^2$ term determines $B_0,C_0$:
\begin{eqnarray}
 B_0 &=& \pm \frac{g^2 \beta_2/2+g^2 a^4+g^2 a^2 b^2/2+ 2 g^2 a^2 r_m^2 + (a^2-b^2)/2}{2g\Xi_a\Xi_b(1+ag+bg)^2 \sqrt{\delta}},\nonumber \\ \qquad C_0 &=& \pm \frac{g^2 \beta_2/2 + g^2 b^4+g^2 a^2 b^2/2 + 2g^2 b^2 r_m^2 - (a^2-b^2)/2}{2g\Xi_a\Xi_b(1+ag+bg)^2 \sqrt{\delta}},
\end{eqnarray}
where $\delta$ is defined by~(\ref{noCTC}). $A_0$ is now determined by (\ref{eqn:hor2}). Finally, vanishing of the coefficient of the $1/R$ term in $g_{RR}$ determines $A_1$:
\bea
 A_1 &=& \frac{g^2 (C_0 Q_0 - B_0 P_0)}{a^2-b^2} + \frac{g^2 (a^2 -b^2)}{8(B_0 - C_0) (1+ag+bg)^4} \nonumber \\ &-& \frac{(a+b)}{2(a^2-b^2)(B_0 - C_0)} \left[B_0^2 (1+bg)(3 b^2 g - a + b+abg) -
 2B_0 C_0 (a^2+b^2) g (2 + ag + bg) \right. \nonumber \\
 &{}& \qquad \qquad \qquad \qquad \qquad \qquad + \left. C_0^2 (1+ag)(3a^2g +a -b+abg) \right].
\eea
We now have
\be
 g_{RR} = {\cal O}(1).
\ee
Using the solutions for $B_0,C_0$ determines
\be
 g_{vR} = \pm \frac{F(\rho^2)^{1/3}}{2g\Xi_a\Xi_b\sqrt{\delta}} + {\cal O}(R).
\ee
The metric and its inverse are now analytic at $R=0$ and can therefore be extended into a new region $R < 0$. The surface $R=0$ is a Killing horizon. To see this, consider the supersymmetric Killing field $V = \partial/\partial v$, which is null at $R=0$. The above results imply
\be
 V_\mu dx^\mu|_{R=0} = \pm \left[ \frac{F(\rho^2)^{1/3}}{2g\Xi_a\Xi_b\sqrt{\delta}} dR \right]_{R=0}.
\ee
This reveals that the $R=0$ is a null hypersurface with normal $V$, i.e., a Killing horizon of $V$. The upper choice of sign corresponds to a future horizon and the lower choice to a past horizon.

Equation (\ref{eqn:Vstatic}) determines the angular velocities of the horizon with respect to the static frame at infinity:
\be
 \Omega_\phi = \Omega_\psi = g.
\ee
The geometry of a spatial cross-section of the event horizon is determined by setting $v={\rm constant}$ and $R=0$, or equivalently $t={\rm constant}$ and $r \rightarrow 0$. This gives a deformed $S^3$ with area
\be
\mathcal{A}_H = 2\pi^2\sqrt{(1+g^2\alpha_1)\alpha_3 -
  \frac{g^2\alpha_2^2}{4} -
  \frac{(A-B)^2}{ABg^6}\left(1+g^2\alpha_{1}+g^4\alpha_2 +
  g^6\alpha_3\right)}.
\ee
This reduces correctly to the results of \cite{gr2,CCLP2,CCLP} in the appropriate limits.

\section{Solutions of the $U(1)^3$ theory}

In this section we shall consider the special case of ${\cal N}=1$ $U(1)^3$ gauged supergravity. This theory is of particular interest because its asymptotically anti-de Sitter solutions can be oxidized to asymptotically $AdS_5 \times S^5$ solutions of type IIB supergravity \cite{11authors}. We shall present a four-parameter supersymmetric black hole solution and list its properties. This section can be read independently of the previous one. However, it is necessary to refer to the previous section for derivations of our results.

The $U(1)^3$ gauged supergravity theory of interest is described in \cite{11authors} (and in section \ref{sec:theory} above). The bosonic sector consists of the graviton, three vectors $A^I$ (denoted $A_i$ in \cite{11authors}) and three real scalars $X^I$ (denoted $X_i$ in \cite{11authors}) subject to the constraint
\be
 X^1 X^2 X^3 = 1.
\ee
We use coordinates $(t,r,\theta,\phi,\psi)$ where $r > 0$ will correspond to the exterior of the black hole, $0 \le \theta \le \pi/2$ and $\phi,\psi$ have period $2\pi$. Surfaces of constant $t$ and $r$ have topology $S^3$.
The space-time metric is:
\be
 ds^2 = -\left(H_1 H_2 H_3 \right)^{-2/3} \left(dt + \omega_\phi d\phi + \omega_\psi d\psi \right)^2 + \left(H_1 H_2     H_3 \right)^{1/3} h_{mn} dx^m dx^n,
\ee
where
\be
 H_I = 1 + \frac{\sqrt{\Xi_a \Xi_b}(1 + g^2\mu_I) - \Xi_a \cos^2 \theta - \Xi_b \sin^2 \theta}{g^2r^2},
\ee
\bea
 h_{mn} dx^m dx^n &=& r^2 \left\{
\frac{dr^2}{\Delta_r} +\frac{d\theta^2}{\Delta_{\theta}} +
\frac{\cos^2\theta}{\Xi_b^2} \left[ \Xi_b + \cos^2 \theta \left(\rho^2 g^2 + 2(1+bg)(a+b)g \right) \right]
d\psi^2 \right. \nonumber \\ &+&
\frac{\sin^2\theta}{\Xi_a^2}\left[
\Xi_a + \sin^2 \theta \left(\rho^2 g^2 + 2(1+ag)(a+b) g \right)
\right]d\phi^2 \nonumber \\  &&\left. +
\frac{2\sin^2\theta \cos^2\theta}{\Xi_a\Xi_b}\left[ \rho^2 g^2
+2(a+b) g+(a+b)^2g^2\right]d\psi d\phi \right\},
\eea
\bea
 \Delta_r &=& r^2[g^2r^2+(1+ag+bg)^2], \qquad \Delta_\theta = 1 - a^2 g^2 \cos^2\theta - b^2 g^2
\sin^2\theta, \nonumber \\
 \Xi_a &=& 1-a^2 g^2, \qquad \Xi_b = 1-b^2 g^2, \qquad \rho^2 = r^2 + a^2 \cos^2\theta + b^2 \sin^2\theta,
\eea
\begin{eqnarray}
 \omega_\psi &=& - \frac{g \cos^2 \theta}{r^2 \Xi_b} \left[ \rho^4 + (2r_m^2 + b^2) \rho^2 + \frac{1}{2} \left(\beta_2-a^2 b^2 + g^{-2} (a^2-b^2)\right) \right], \nonumber \\
 \omega_\phi &=& - \frac{g \sin^2 \theta}{r^2 \Xi_a} \left[ \rho^4 + (2r_m^2 + a^2) \rho^2 + \frac{1}{2} \left(\beta_2-a^2 b^2 - g^{-2} (a^2-b^2)\right) \right],
\end{eqnarray} and
\begin{equation} r_m^2=g^{-1} (a + b) + ab
\end{equation}
\begin{equation}\beta_2 = \Xi_a\Xi_b(\mu_1\mu_2+\mu_1\mu_3+\mu_2\mu_3) -\frac{2\sqrt{\Xi_a\Xi_b}\left(1-\sqrt{\Xi_a\Xi_b} \right)}{g^2}(\mu_1+\mu_2+\mu_3) + \frac{3\left(1-\sqrt{\Xi_a\Xi_b}\right)^2}{g^4}
\end{equation}
The scalars are
\be
 X^I = \frac{\left(H_1 H_2 H_3 \right)^{1/3}}{H_I}.
\ee
The vectors are:
\be
 A^I = H_I^{-1}(dt + \omega_{\psi} d\psi + \omega_{\phi} d\phi) + U^I_{\psi} d\psi + U^I_{\phi} d\phi
\ee where
\bea
U^I_{\psi} &=& \frac{g\cos^2\theta}{\Xi_b}\left(\rho^2 + 2r_m^2 + b^2  -\sqrt{\Xi_a\Xi_b}\mu_I + g^{-2}\left(1-\sqrt{\Xi_a\Xi_b} \right) \right) \nonumber \\
U^I_{\phi} &=& \frac{g\sin^2\theta}{\Xi_a}\left(\rho^2 + 2r_m^2 + a^2  -\sqrt{\Xi_a\Xi_b}\mu_I + g^{-2}\left(1-\sqrt{\Xi_a\Xi_b} \right) \right)
\eea
The solution depends on five parameters:\footnote{
We emphasize that the parameters $a,b$ are {\it not} the same as in~\cite{CCLP2,CCLP}. See section \ref{sec:base} for details. For comparison with section 2 note that $q_I=\mu_I/3$.}
$\mu_1,\mu_2,\mu_3,a,b$ where $g^{-1} >a,b \ge 0$. Only four parameters are independent because of the constraint
\be
 \mu_1 + \mu_2 + \mu_3 = \frac{1}{\sqrt{\Xi_a \Xi_b}} \left[ 2r_m^2 + 3 g^{-2} \left( 1 - \sqrt{\Xi_a \Xi_b} \right) \right].
\ee
Regularity of the scalars for $r \ge 0$ requires that
\be
 g^2 \mu_I > \sqrt{\frac{\Xi_b}{\Xi_a}} -1  \ge 0,
\ee
when $a \ge b$. If $a<b$ then the same expression applies with $a,b$ interchanged.

This solution is expressed in co-rotating coordinates. A coordinate transformation $t = \bar{t}$, $\phi = \bar{\phi} - g \bar{t}$, $\psi = \bar{\psi} - g \bar{t}$ is required to bring it to a manifestly asymptotically anti-de Sitter form as $r \rightarrow \infty$.
\par
The electric charges are
\bea
Q_1 &=& \frac{\pi}{4G} \left[\mu_1 +\frac{g^2}{2}\left(\mu_1\mu_2+\mu_1\mu_3-\mu_2\mu_3 \right) \right] \nonumber \\
Q_2 &=& \frac{\pi}{4G} \left[\mu_2 +\frac{g^2}{2}\left(\mu_2\mu_3+\mu_2\mu_1-\mu_1\mu_3 \right) \right] \nonumber \\
Q_3 &=& \frac{\pi}{4G} \left[\mu_3
+\frac{g^2}{2}\left(\mu_3\mu_2+\mu_1\mu_3-\mu_1\mu_2 \right)
\right]. \eea Let
\begin{equation}
\mathcal{J} \equiv (1+g^2\mu_1)(1+g^2\mu_2)(1+g^2\mu_3).
\end{equation} The angular momenta are given by
\bea
J_{\phi} &=& \frac{\pi}{4G}\left( \frac{1}{2}g(\mu_1\mu_2+\mu_2\mu_3+\mu_1\mu_3)+ g^3\mu_1\mu_2\mu_3
+g^{-3}\left(\sqrt{\frac{\Xi_b}{\Xi_a}}-1\right)\mathcal{J}\right),
\\
J_{\psi} &=& \frac{\pi}{4G}\left( \frac{1}{2}g(\mu_1\mu_2+\mu_1\mu_3+\mu_2\mu_3)+ g^3\mu_1\mu_2\mu_3
+g^{-3}\left(\sqrt{\frac{\Xi_a}{\Xi_b}}-1\right)\mathcal{J} \right).
\eea The mass is deduced from the BPS equality (\ref{eqn:bps}):
\begin{equation}
M= \frac{\pi}{4G} \left(\mu_1+\mu_2+\mu_3 + \frac{3}{2}g^2(\mu_1\mu_2+\mu_1\mu_3+\mu_2\mu_3) + 2g^4\mu_1\mu_2\mu_3 +\frac{(\sqrt{\Xi_a}-\sqrt{\Xi_b})^2}{g^2\sqrt{\Xi_a\Xi_b}}\mathcal{J} \right).
\end{equation}
The solution has an event horizon at $r=0$. The angular velocities of the event horizon with respect to the static frame at infinity are
\begin{equation}
 \Omega_\psi = \Omega_\phi = g
\end{equation}
Spatial cross sections of the event horizon have the geometry of a deformed $S^3$ with area
\be\label{AreaUcubed}
\mathcal{A}_H = 2\pi^2\sqrt{[1+g^2(\mu_1+\mu_2+\mu_3)]\mu_1\mu_2\mu_3 -
  \frac{g^2(\mu_1\mu_2+\mu_1\mu_3+\mu_2\mu_3)^2}{4} - \frac{(\sqrt{\Xi_a}-\sqrt{\Xi_b})^2}{g^6\sqrt{\Xi_a\Xi_b}}\mathcal{J}}.
\ee
The expression within the square root must be positive, otherwise there are closed causal curves near $r=0$. Note that ${\cal J}>0$ so, for given electric charges, the horizon area is maximized and the mass minimized when $a=b$, i.e., when $J_\phi=J_\psi$.

\begin{center} {\bf Acknowledgments} \end{center}

We thank Jan Gutowski for useful conversations. HKK would like to
thank St. John's College, Cambridge for financial support. JL would
like to thank St. John's College, Oxford for financial support and
DAMTP, Cambridge for hospitality. HSR is a Royal Society University
Research Fellow. HKK and JL would also like to thank the Particle
Theory group at Nottingham for hospitality. Part of this work was
done whilst HSR was a participant in the the Newton Institute
programme ``Global Problems in Mathematical Relativity" so he thanks
the Institute for hospitality and financial support.

\end{document}